\documentclass[preprint,authoryear,12pt]{elsarticle}

\usepackage{amssymb}
\usepackage{natbib}
\usepackage{amsmath, bm}
\usepackage{graphicx}
\usepackage{lineno} 
\usepackage{float}
 \usepackage{setspace}
\usepackage[nodots]{numcompress}
\usepackage[top=1in, bottom=1in, left=1.25in, right=1.25in]{geometry}
\usepackage[dvipsnames]{xcolor}

\usepackage{algorithm}
\usepackage{algorithmic}

\usepackage{caption}
\usepackage[skip=5pt]{caption}
\usepackage[colorlinks]{hyperref}
\usepackage[nameinlink]{cleveref}

\Crefname{figure}{Fig.}{Figs.}
\begin{document}

\begin{spacing}{1.25}
\begin{frontmatter}

\title{\textbf{Data-driven modelling of nonlinear spatio-temporal fluid flows using a deep convolutional generative adversarial network}}

\author[label1]{M. Cheng}

\author[label1]{F. Fang \corref{cor1}}
\address[label1]{Applied Modelling and Computation Group, Department of Earth Science and Engineering, Imperial College London, SW7 2BP, UK}
\address[label2]{Department of Scientific Computing, Florida State University, Tallahassee, FL, 32306-4120, USA}

\cortext[cor1]{Corresponding author}
\ead{f.fang@imperial.ac.uk}

\author[label1]{C.C. Pain}

\author[label2]{I.M. Navon}

\begin{abstract}
Deep learning techniques for improving fluid flow modelling have gained significant attention in recent years. Advanced deep learning techniques achieve great progress in rapidly predicting fluid flows without prior knowledge of the underlying physical relationships. However, most of existing researches focused mainly on either sequence learning or spatial learning, rarely on both spatial and temporal dynamics of fluid flows \citep{reichstein2019deep}. In this work, an Artificial Intelligence (AI) fluid model based on a general deep convolutional generative adversarial network (DCGAN) has been developed for predicting spatio-temporal flow distributions. In deep convolutional networks, the high-dimensional flows can be converted into the low-dimensional “latent” representations. The complex features of flow dynamics can be captured by the adversarial networks. The above DCGAN fluid model enables us to provide reasonable predictive accuracy of flow fields while maintaining a high computational efficiency. The performance of the DCGAN is illustrated for two test cases of Hokkaido tsunami with different incoming waves along the coastal line. It is demonstrated that the results from the DCGAN are comparable with those from the original high fidelity model (Fluidity). The spatio-temporal flow features have been represented as the flow evolves, especially, the wave phases and flow peaks can be captured accurately. In addition, the results illustrate that the online CPU cost is reduced by five orders of magnitude compared to the original high fidelity model simulations. The promising results show that the DCGAN can provide rapid and reliable spatio-temporal prediction for nonlinear fluid flows. 
\end{abstract}

\begin{keyword}
Deep convolutional GAN\sep nonlinear fluid flow\sep data-driven modelling\sep deep learning
\end{keyword}

\end{frontmatter}

\section{Introduction}
\label{sec1} 
Numerical simulations of fluid dynamics have been indispensable in many applications relevant to physics and engineering. For improving predictive capability, numerical algorithms have become increasingly sophisticated and complicated by using more spatial and temporal resolution of datasets. This often leads to very large-scale models and unavoidably raises computational challenges. In this work, an Artificial Intelligence (AI) fluid flow model is presented for rapidly providing spatio-temporal prediction of nonlinear fluid flows while avoiding the excessive computational cost.

Recently, deep learning has gained popularity in rapidly predicting nonlinear fluid flows \citep{kalteh2013monthly,kasiviswanathan2016potential,kisi2012forecasting,lohani2014improving, yang2017developing, wang2018model, Brunton2019MachineLF}. Deep learning techniques are capable to handle the large-scale data from the dynamic and nonlinear systems, and provide accurate nonlinear fluid flow prediction without prior knowledge of underlying physical relationships. Various deep learning techniques have been applied to nonlinear fluid flow prediction, for example, modifying the model parameters by Bayesian deep neural networks to enhance the flow prediction in the turbulence model \citep{geneva2019quantifying}, obtaining the time-series streamflow prediction by Long Short-Term Memory (LSTM) networks \citep{hu2019rapid}, and simulating unsteady wake flows by convolutional neural network (CNN) -based deep learning \citep{miyanawala2017efficient}. 

More recently, Generative Adversarial Networks (GANs) have been introduced to reconstruction of fluid dynamic features \citep{farimani2017deep, xie2018tempogan}.  GANs represent the most recent progress in deep learning techniques \citep{goodfellow2016nips, laloy2017efficient, strofer2018data} and have been attractive for producing high-resolution samples (e.g. images) \citep{ledig2017photo, xie2018tempogan, creswell2018generative}. The main specificity of GANs is the competition game between the generator and discriminator modules. The generator consists of deep neural networks, which can produce high-dimensional distributions of data to “fool” the discriminator. The discriminator is responsible for distinguishing the real data from the generated data. Nowadays applications of GANs focus mostly on images \citep{reed2016generative, li2016precomputed, isola2017image}, rarely on fluid flow problems \citep{farimani2017deep, xie2018tempogan}. \citet{farimani2017deep} first introduced the conditional generative adversarial networks (cGAN) to simulate steady state heat conduction and incompressible fluid flow. This study proves that GANs can be effective in exhibiting complex nonlinear fluid structures. \citet{xie2018tempogan} introduced a temporal discriminator to GANs for solving the super-resolution problem for smoke flows. 

In this work, we have developed a deep convolutional generative adversarial network (DCGAN) \citep{radford2015unsupervised} for nonlinear unsteady fluid flow problems. This is the first attempt to use the DCGAN for predictive analysis in fluid flow problems. As is well known, both spatially and temporally flow dynamic properties are important in prediction. Classic machine learning approaches exploit dependencies in time or space, rarely both spatio-temporal dependencies \citep{reichstein2019deep}. Here, the DCGAN is first used to extract the spatio-temporal coherence of the original high-dimensional fluid fields. The high-dimensional nonlinear fluid flow features can be converted into low-dimensional “latent” representations. With discriminating the real fluid fields and their realizations in the low-dimensional latent space, the training process herein is fast. This alleviates the problem of large data-driven modelling. The way of dimensionality reduction in the deep convolutional networks is similar to that of Galerkin reduced order model (ROM) \citep {fang2013non, fang2014reduced, xiao2019machine}, Variational Autoencoder (VAE) \citep{ gonzalez2018learning} and cluster-based reduced order model (CROM) \citep{kaiser2014cluster, nair2019cluster, osth2015cluster}. These ROMs are able to compress spatial-temporal flow snapshots into a small number of low-dimensional states for dynamic modelling, which is also a strategy to reduce computational requirements of ROMs \citep{kaiser2014cluster}. Nevertheless, the prominent difference between the DCGAN and existing ROMs lies in the fact that the CROM or POD based ROMs result in linear combinations of the original snapshots \citep{kaiser2014cluster, xiao2016non}, while the deep convolutional neural networks are able to capture the nonlinear flow transients since the effective basis functions are non-linear \citep{miyanawala2017efficient}. In addition, GAN also has the key advantage that, in principle, any point in the latent space of inputs will be associated with a realistic looking solution.

The DCGAN is a general and efficient numerical tool for rapidly predicting nonlinear fluid flows for given inputs. In this work, we have successfully applied the DCGAN to a tsunami case in the southwestern Hokkaido Island, Japan. The performance of the DCGAN has been evaluated by comparing the results obtained with those from an unstructured mesh finite element model (Fluidity) \citep{pain2001tetrahedral}.

The structure of this article is organized as follows. \Cref{sec2} presents the governing equation for nonlinear fluid flows in dynamic systems. The architecture of the DCGAN is briefly introduced in \cref{sec3}. The detailed process for the large data-driven modelling is described in \cref{sec4}. \Cref{sec5} illustrates the capability of the DCGAN in the tsunami case, which entails the snapshot-based datasets generation with the different wave characteristics and the DCGAN training. Results of cases and the model efficiency are then presented in \cref{sec6}. Conclusions are drawn in \cref{sec7}.

\section{Governing equations for nonlinear fluid flow prediction problems}
\label{sec2}
Nonlinear fluid flows are usually simulated by solving the complex non-linear equations (for example, Navier-Stokes (NS) equations, Lorenz equations and shallow water equations) in the computational fluid dynamic models. In a general way, the governing equation of dynamic flows can be expressed as:
\begin{equation}
\label{eq:4}
\frac{{d\phi}}{{dt}} = \frac{d}{{dt}}f({\mu}, s, {\bf x}, t),
\end{equation}
where ${\phi}$ is the state variable to be predicted (for example, water depth and velocity), ${\mu}$ represents the inputs (for example, initial and boundary conditions, modelling parameters) , ${\bf x}=(x,y,z)$ denotes the spatial coordinate system, $s$ is the source term, and $t$ is the time.

This work focuses on the spatio-temporal fluid simulations that are driven by the inputs (for example, initial and boundary conditions, modelling parameters). The nonlinear relationship between the fluid flow solutions $\phi$ and the inputs $\mu$ in Eq.\eqref{eq:4} can be re-written in a general way:
\begin{equation}
\label{relationship}
{\phi} = {\mathcal{P}}({\mu}),
\end{equation}
where ${\mathcal {P}}$ represents a dynamic model. In this paper, the dynamic model ${\mathcal {P}}$ will be represented by the DCGAN introduced in \cref{sec4}. Based on a series of inputs ${\mu}$, the main objective of this work is to develop a new DCGAN tool for simulating the spatio-temporal distribution of nonlinear fluid flows ${\phi}$. 

\section{Adversarial network architecture}
\label{sec3}
\begin{figure}[H]
\centering
\includegraphics[width=1.0\textwidth]{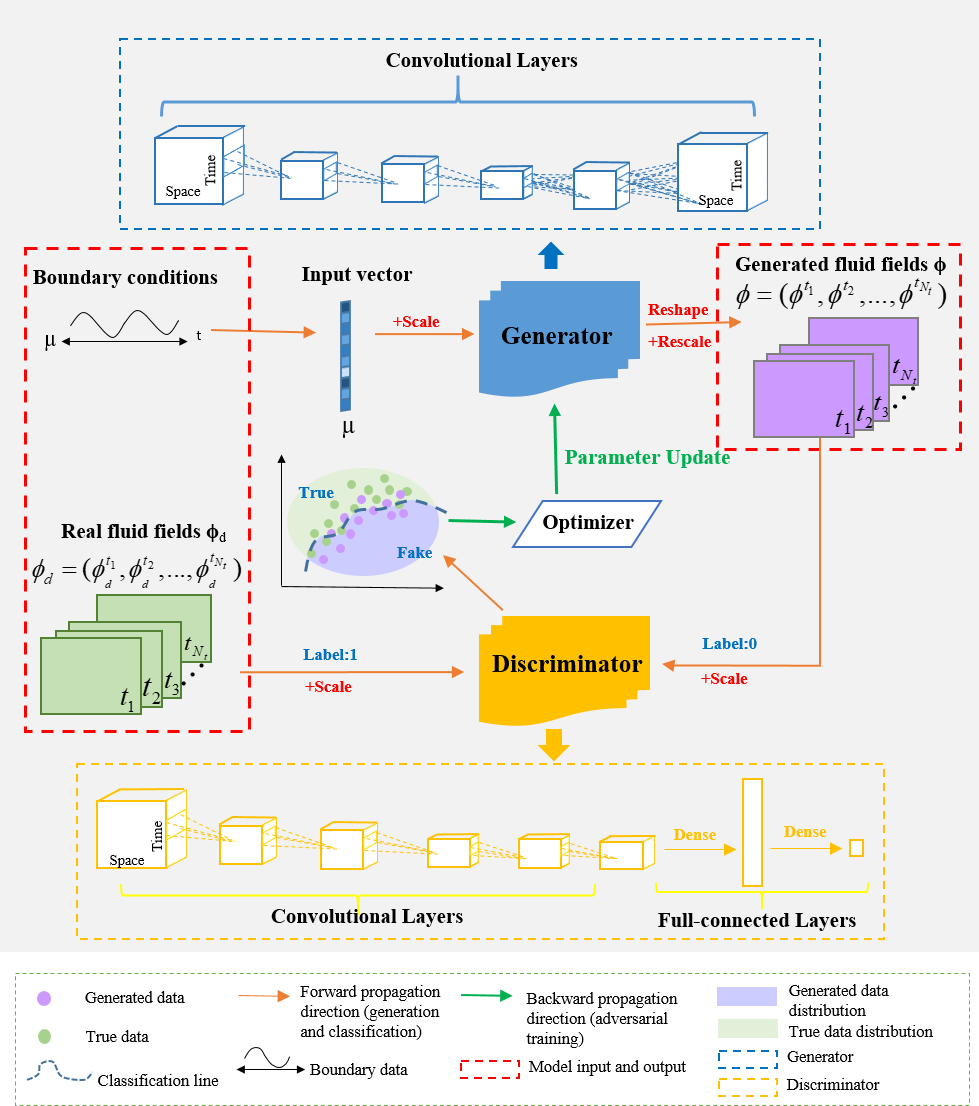}
\caption{Deep convolutional generative adversarial network architecture. The $\phi$ and ${\phi_d}$ can be represented by fluid flow solutions (e.g. water depth).}
\label{GAN}
\end{figure}

A generative adversarial network comprises a generator module (G) and a discriminator module (D), as shown in \Cref{GAN}. The role of the generator is to map a random dataset ${\mu}$ (taking as inputs) into a desired output dataset ${\phi}$. In traditional GANs, the random dataset $\mu$ is typically a Gaussian or uniform noise ranging from zero to one. The discriminator is used for distinguishing the real datasets ${\phi_d}$ (the targeted outputs) from the generated outputs ${\phi}$. In the discriminator, ${D(\phi_d)=1}$ if the targeted outputs are accepted while ${D(G(\mu)) =0}$ if they are rejected.

In general, the training process in GAN can be considered as a minimization-maximization problem based on a cross entropy loss function ${\bf J}(D, G))$:
\begin{equation}
\label{eq:3}
\mathop {\min }\limits_G \mathop {\max }\limits_D {E_{\phi_d\sim{p_{data}(\phi_d)}}}[\log D(\phi_d)] + {E_{\mu\sim{p_{\mu}(\mu)}}}[\log (1 - D(G(\mu)))],
\end{equation}
where ${p_{\mu}(\mu)}$ is a prior distribution for the random dataset ${\mu}$, and ${p_{data}(\phi_d)}$ is the corresponding probability data distribution for the targeted outputs ${\phi_d}$.

In particular, the min-max process consists of two steps:
\begin{itemize}
\item Given a fixed generator, updating the discriminator by maximizing the discriminator function $\mathop {\max }\limits_{\theta^{(D)}}{{\bf J}^{(D)}} \left( {{\theta ^{\left( D \right)}},{\theta ^{\left( G \right)}}}\right)$ (where ${\theta ^{\left( D \right)}}$ represents the parameters
 in the cost functions of the discriminator, and the function ${\bf J}^{(D)}$ is differentiable.). 
  \begin{equation}
\label{eq:Jd}
{\bf{J}}^{(D)} \left( {{\theta ^{\left( D \right)}},{\theta ^{\left( G \right)}}}\right) = {E_{{\phi_d}\sim{p_{data}(\phi_d)}}}[\log D({\phi_d}, \theta^{(D)})] + {E_{{\mu}\sim{p_\mu(\mu)}}}[\log (1 - D(G({\mu}), \theta^{(G)}, \theta^{(D)})];
\end{equation}
\item Updating the generator by minimizing the generator function, i.e. $\mathop {\min }\limits_{\theta^{(G)}}{\bf J}^{(G)} \left( {{\theta ^{\left( D \right)}},{\theta ^{\left( G \right)}}}\right)$ (where ${\theta ^{\left( G \right)}}$ represents the parameter in the cost functions of the generator, and the function ${\bf J}^{(G)}$ is differentiable.)
  \begin{equation}
\label{eq:1}
{\textbf{J}^{(G)}}\left( {{\theta ^{\left( D \right)}},{\theta ^{\left( G \right)}}}\right) = {E_{\mu\sim{p_\mu(\mu)}}}[\log (1 - D(G(\mu)))].
\end{equation}
\end{itemize}

Following \cite{goodfellow2014generative}, the min-max game has a global optimum for ${{p_g}(\phi) = {p_{data}(\phi_d)}}$ (where ${{p_g}}$ is a probability distribution of the generated output ${\phi}$ when ${\mu\sim{p_{\mu}(\mu)}}$), if $D$ and $G$ have enough capacity. The optimal discriminator for the ${\textbf {J}^{(G)} (G,D)}$ can be characterized by: 
\begin{equation}
{D^{*}}(\phi_d) = \frac{{{p_{data}}(\phi_d)}}{{{p_{data}}(\phi_d) + {p_g}(\phi)}}
\end{equation}

The minimax game in \eqref {eq:3} can be rewritten as 
\begin{equation}
\mathop {\max }\limits_D \textbf{J}^{(G)}(G,D) = {E_{\phi_d \sim {p_{data}}}}[\log \frac{{{p_{data}}(\phi_d)}}{{{p_{data}}(\phi_d) + {p_g}(\phi)}}] + {E_{\phi \sim {p_g}}}[\log \frac{{{p_g}(\phi)}}{{{p_{data}}(\phi_d) + {p_g}(\phi)}}]
\end{equation}

When the ${{D^ * }(\phi_d) = \frac{1}{2}}$, the generator can use any random dataset ${\mu}$ to synthesis the generated outputs ${\phi}$, and the discriminator loses the ability to distinguish between the real dataset ${\phi_d}$ and generated ones ${\phi}$.

Gradient-based algorithms can be used for optimization of parameters $\theta^{(G)}, \theta^{(D)}$ in the generator and discriminator. For details see sections below.

\section{Deep convolutional GAN for spatio-temporal data-driven  modelling}
\label{sec4}
Details about model development for spatio-temporal fluid flow modelling, please see full paper.

\section{Numerical examples}
\label{sec5}
In this section, two flooding examples are presented to illustrate the capabilities of the DCGAN in resolving nonlinear fluid flow problem governed by the NS equations. In these examples, the inputs are a series of incoming waves along the coastal line while the corresponding outputs are the flooding solutions (e.g. water depth).

To demonstrate the predictive capability of the DCGAN for dynamic fluid flows, it has been applied to a case of flooding in Hokkaido, Japan. The flooding happened in 1993, caused by the Hokkaido Nansei-Oki earthquake offshore of southwestern Hokkaido Island, Japan. Incoming waves along the coastal line have been identified as an important driver for flood hazards in coastal areas. The increased demand to improve flood prediction requires better understanding of uncertainties associated with flood hazards due to the variability of waves \citep{hu2018unstructured, hu2019numerical}. In our flooding study cases, the characteristics of waves including the wave phase and wave height were considered. The phase variation represents the lags in timing of simulated flows \citep{prakash2014improved, chang2016nonlinear}, and the wave height variation indicates the magnitude of the flow peak \citep{kasiviswanathan2016potential, chang2016nonlinear}. 

\begin{figure}[H]
\centering
\includegraphics[width=0.9\textwidth]{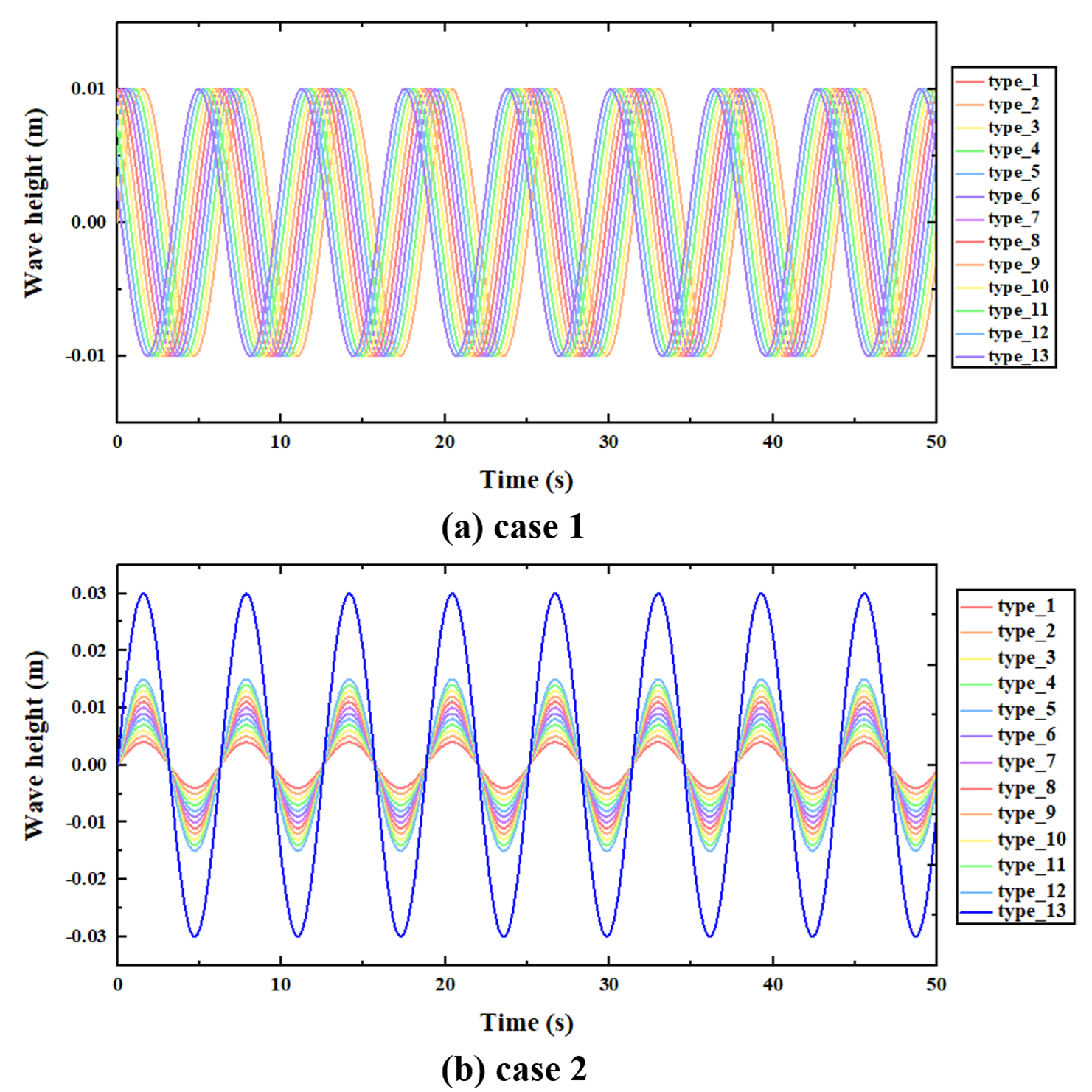}
\caption{Input varied waves for two cases: (a) representing the wave phase variation and (b) representing the wave height variation.}
\label{input_waves}
\end{figure}

To investigate the danger caused by the wave run-up and flooding, two series of incoming waves $(H^{(e)}, e = (1,2))$ are given and included in the inputs ${\mu}$ in two study cases, where each series of waves contains 13 types of waves, as shown in \Cref{input_waves}. 

\begin{figure}[H]
\centering
\includegraphics[width=0.8\textwidth]{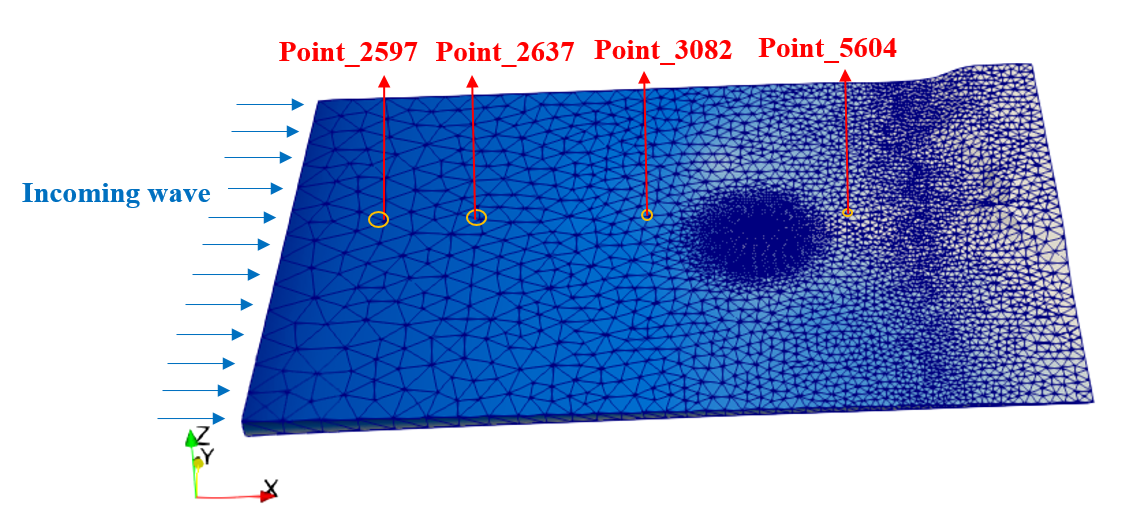}
\caption{Computational domain area and unstructured mesh (the incoming wave is the boundary condition along the costal line, and the simulation starts with a static state).}
\label{geometry}
\end{figure}

\section{Results}
\label{sec6}

\subsection{{Case 1: Wave phase variation}}
\label{subsec1}
In this case, the numerical illustration of the DCGAN was set up with a series of waves where the wave phase varies in the Hokkaido tsunami case. Given a new input ${\mu \in (\vartheta \backslash {\vartheta _{tr}})}$, a comparison between the DCGAN and the original high fidelity model results is carried out. 

\begin{figure}[H]
\centering
\includegraphics[width=1.0\textwidth]{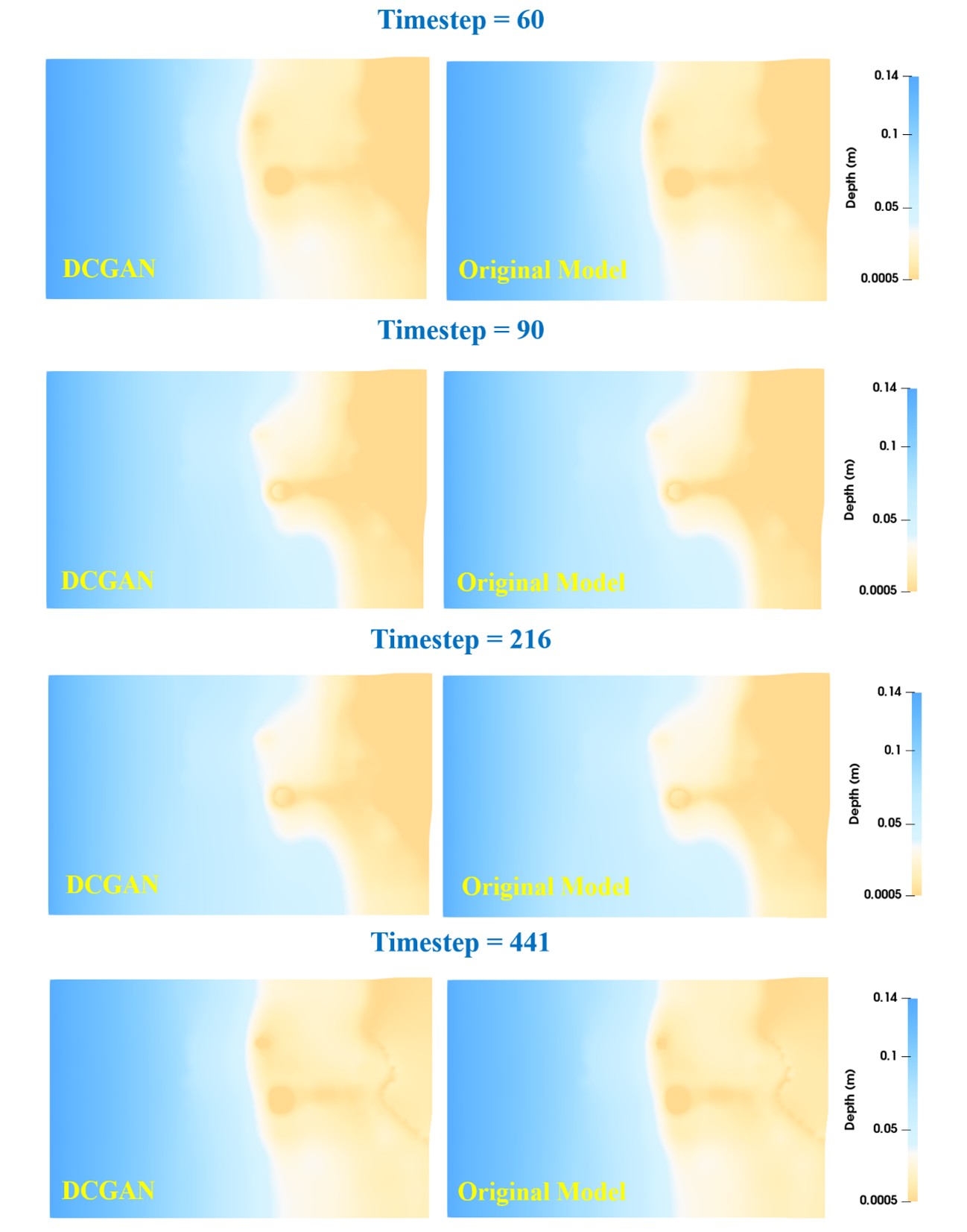}
\caption{Comparison of the spatial distribution of water depth fields obtained from the DCGAN (left) and the original high fidelity model (right) at timesteps $60, 90, 216, 441$.}
\label{fields1}
\end{figure}

\begin{figure}[H]
\centering
\includegraphics[width=1.0\textwidth]{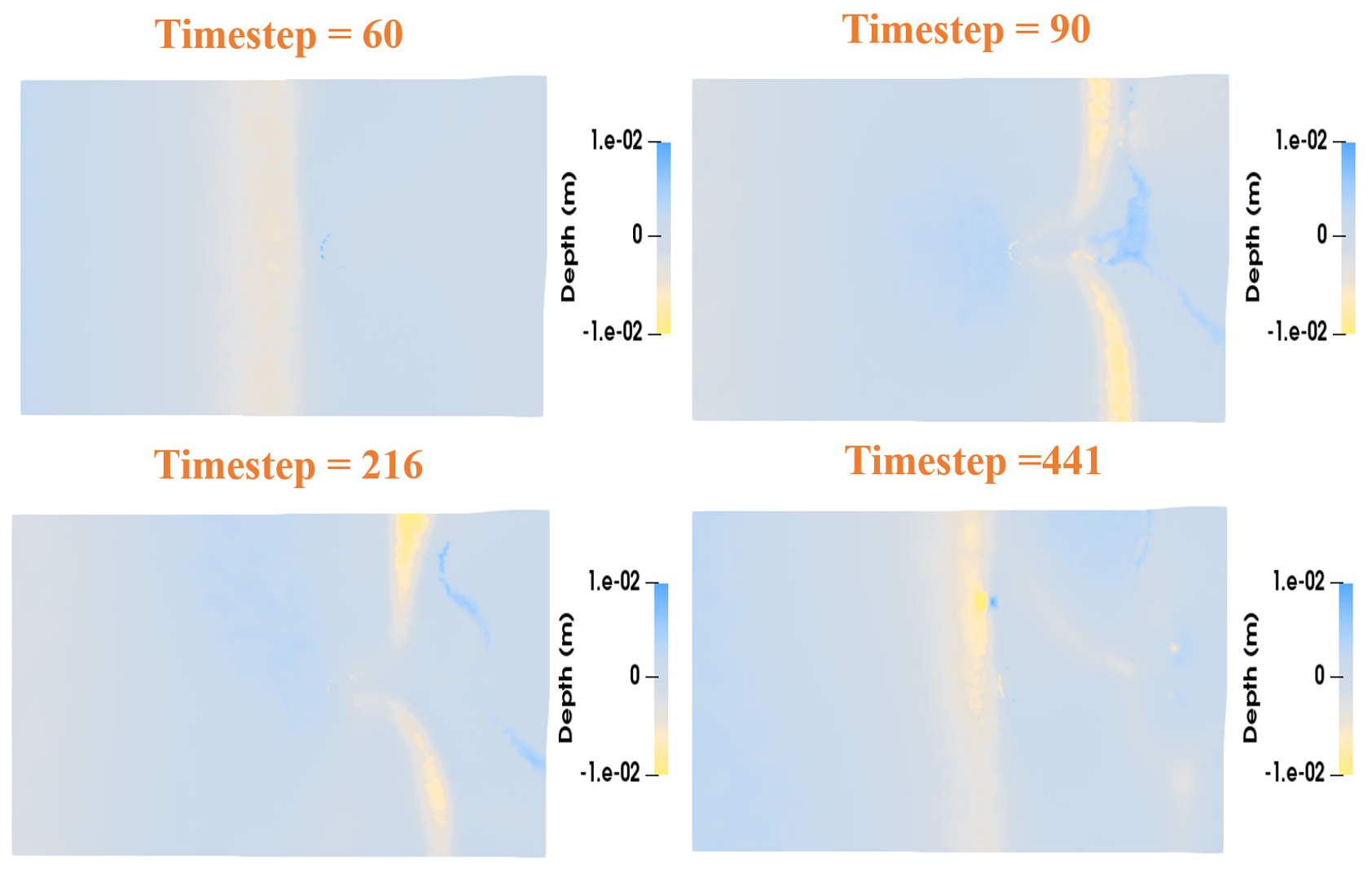}
\caption{Differences of water depth fields between the DCGAN and the original high fidelity model at timesteps $60, 90, 216, 441$.}
\label{errors_fields}
\end{figure}

\Cref{fields1} presents the comparison of water depth obtained from the DCGAN and the original high fidelity model at timesteps $60, 90, 216, 441$. It is worth noting that the results from the DCGAN attain a closer agreement to those from the original high fidelity model at each timestep. The differences of the water depth between the DCGAN and the original high fidelity model are illustrated in \Cref{errors_fields}. The differences are quite small over the whole domain, which suggests the DCGAN is able to obtain reasonable and accurate solutions for nonlinear fluid flows. 

\begin{figure}[H]
\centering
\includegraphics[width=1.0\textwidth]{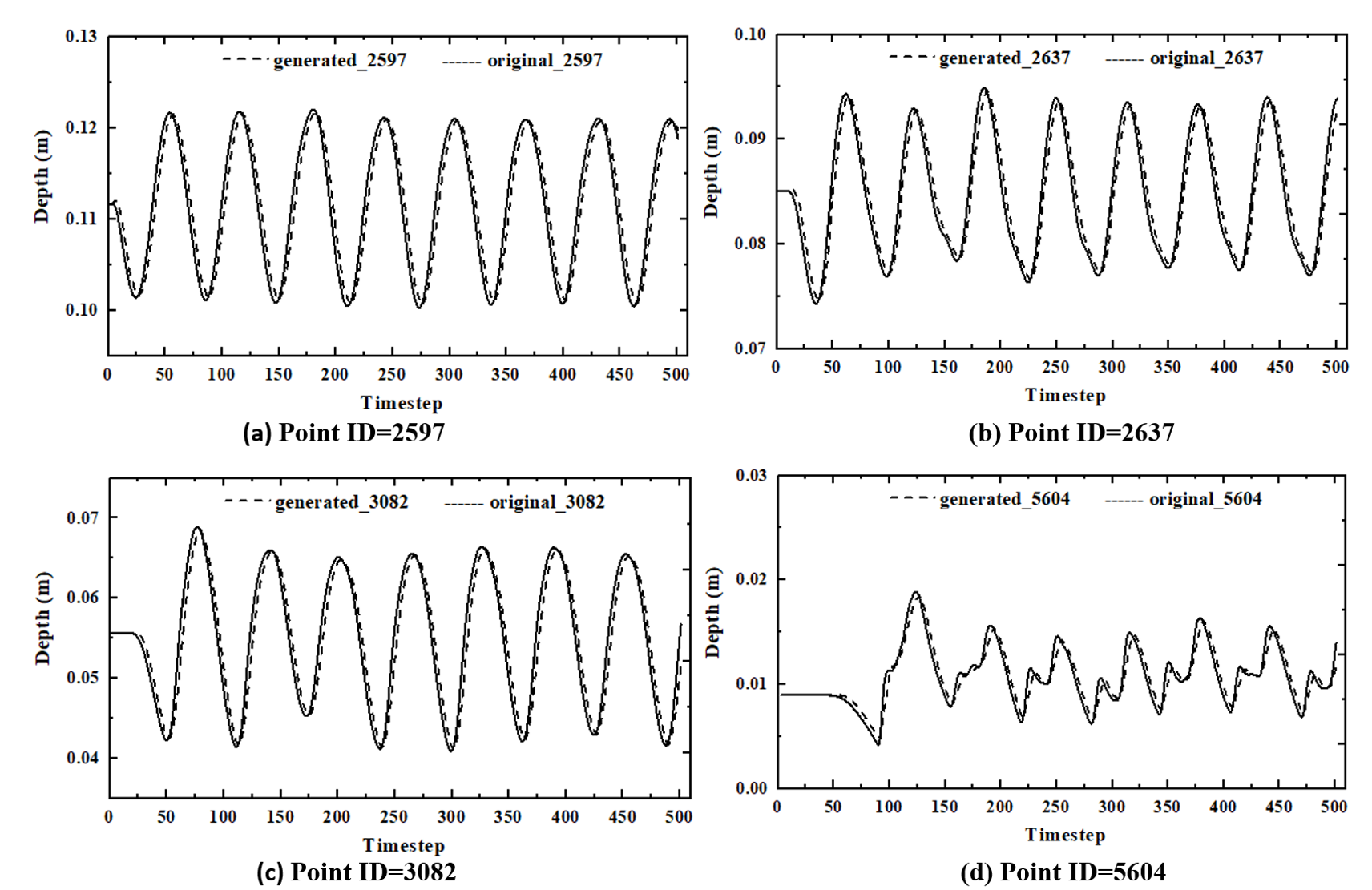}
\caption{Comparison of the temporal variation of water depth at points ID = 2597, 2637, 3082, 5604.}
\label{points}
\end{figure}

As shown in \Cref{points}, the temporal variation of water depth is demonstrated at four points (locations shown in \Cref{geometry}). It can be observed that the curves of water depth from the DCGAN have a slight difference from that of the original high fidelity model. In addition, the arriving time lags at four points can be noticed. The arrival time of flooding at point ID = 5604 is at timestep 50, where it is the farthest from the coastal boundary among the four points. 

\begin{figure}[H]
\centering
\includegraphics[width=0.7\textwidth]{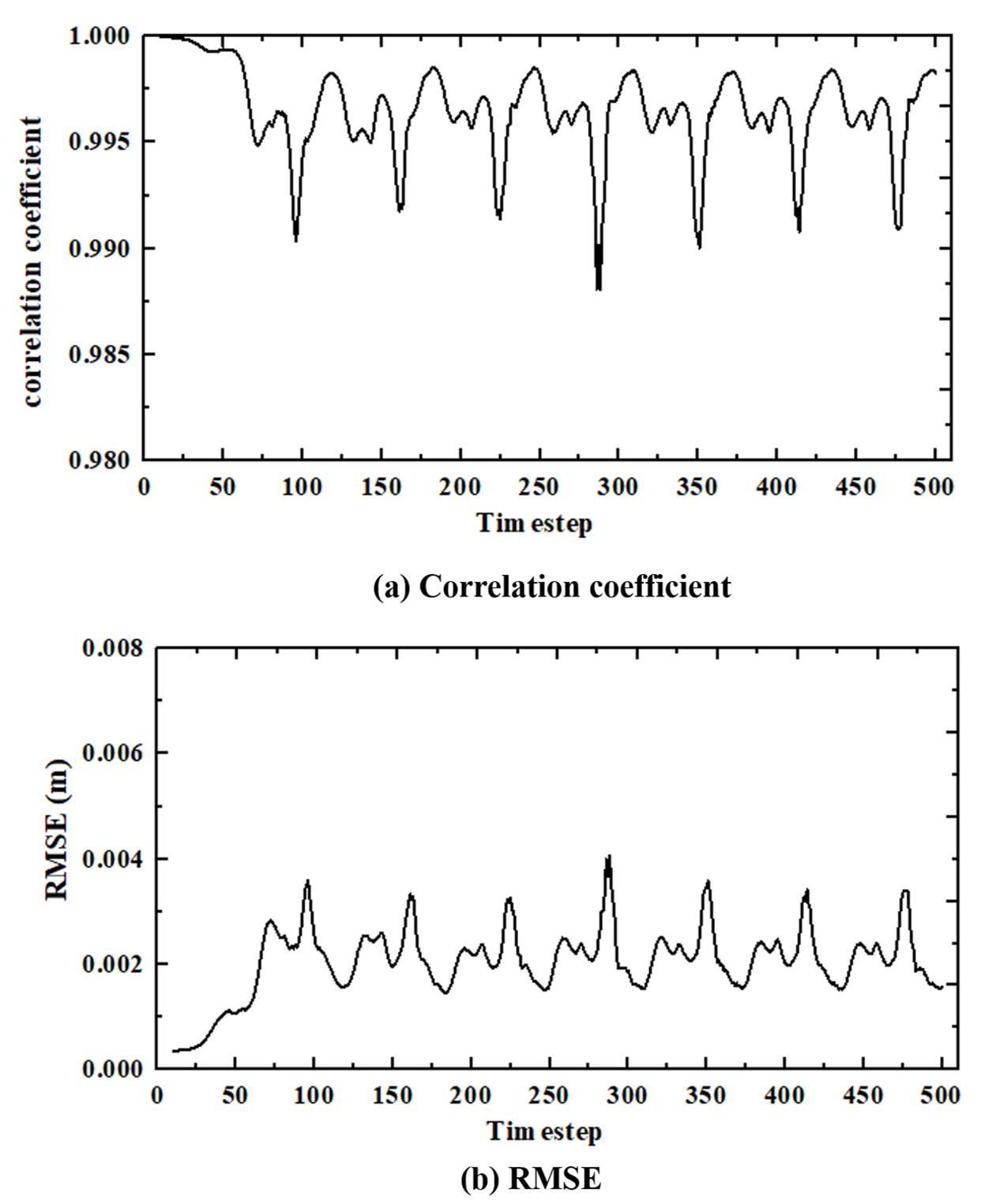}
\caption{The RMSE and correlation coefficients of water depth between the generated fields and original high fidelity fields during whole simulational period.}
\label{RMSE_R}
\end{figure}

To further estimate the performance of the DCGAN, the correlation coefficient and  RMSE are further illustrated in \Cref{RMSE_R}. It is evident that the values of RMSE are between $0.002 m$ to $0.004 m$ while the correlation coefficients are above ${99\%}$. These results demonstrate that the DCGAN performs very well and the predicted fluid fields are in good agreement with the true fluid fields.

\subsection{{Case 2: Wave peak variation}}
\label{subsec2}
\begin{figure}[H]
\centering
\includegraphics[width=1.0\textwidth]{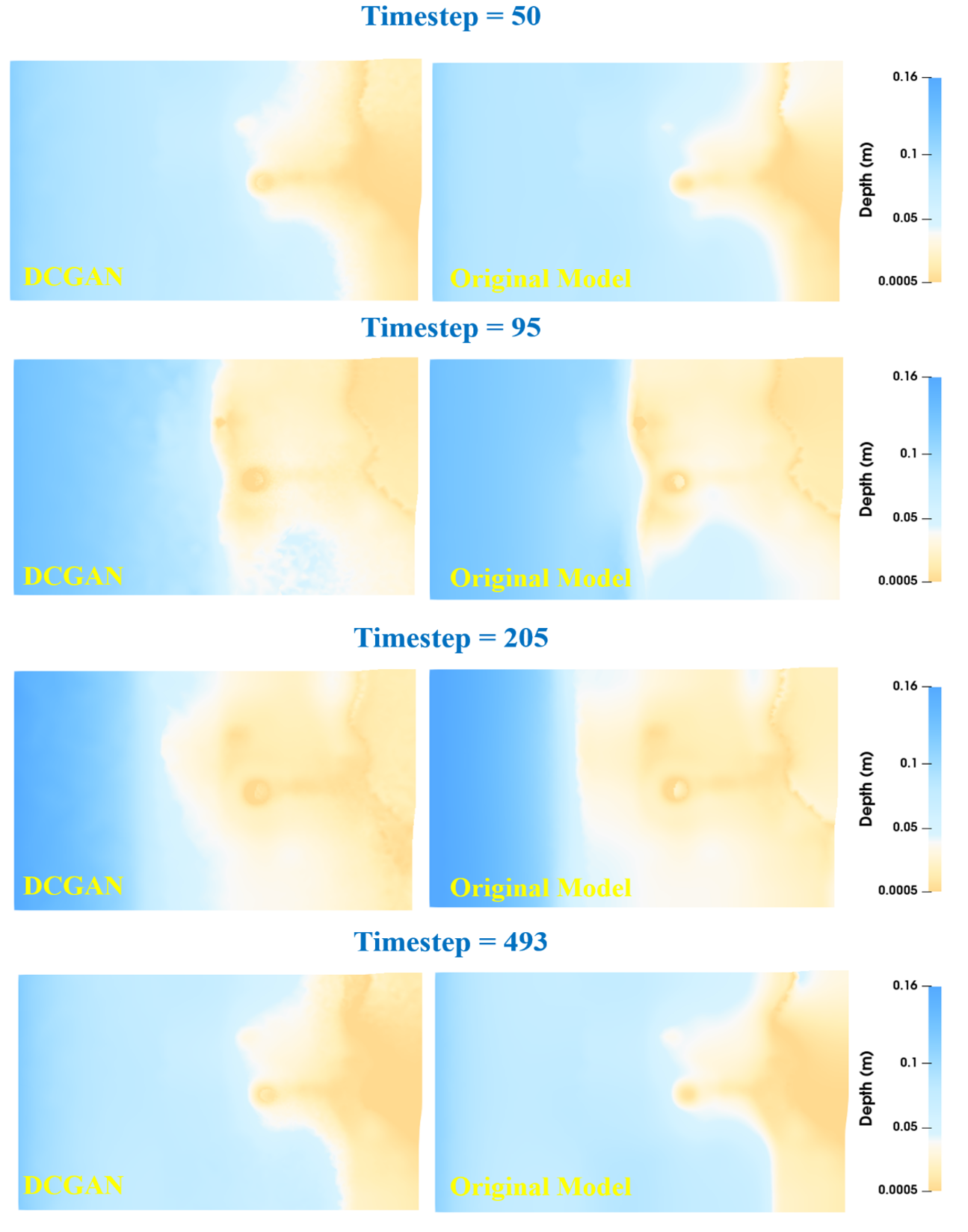}
\caption{Comparison of spatial distribution of water depth fields obtained from the DCGAN (left) and the original high fidelity model (right) at timesteps $50, 95, 205, 493$.}
\label{fields2}
\end{figure}

\begin{figure}[H]
\centering
\includegraphics[width=0.8\textwidth]{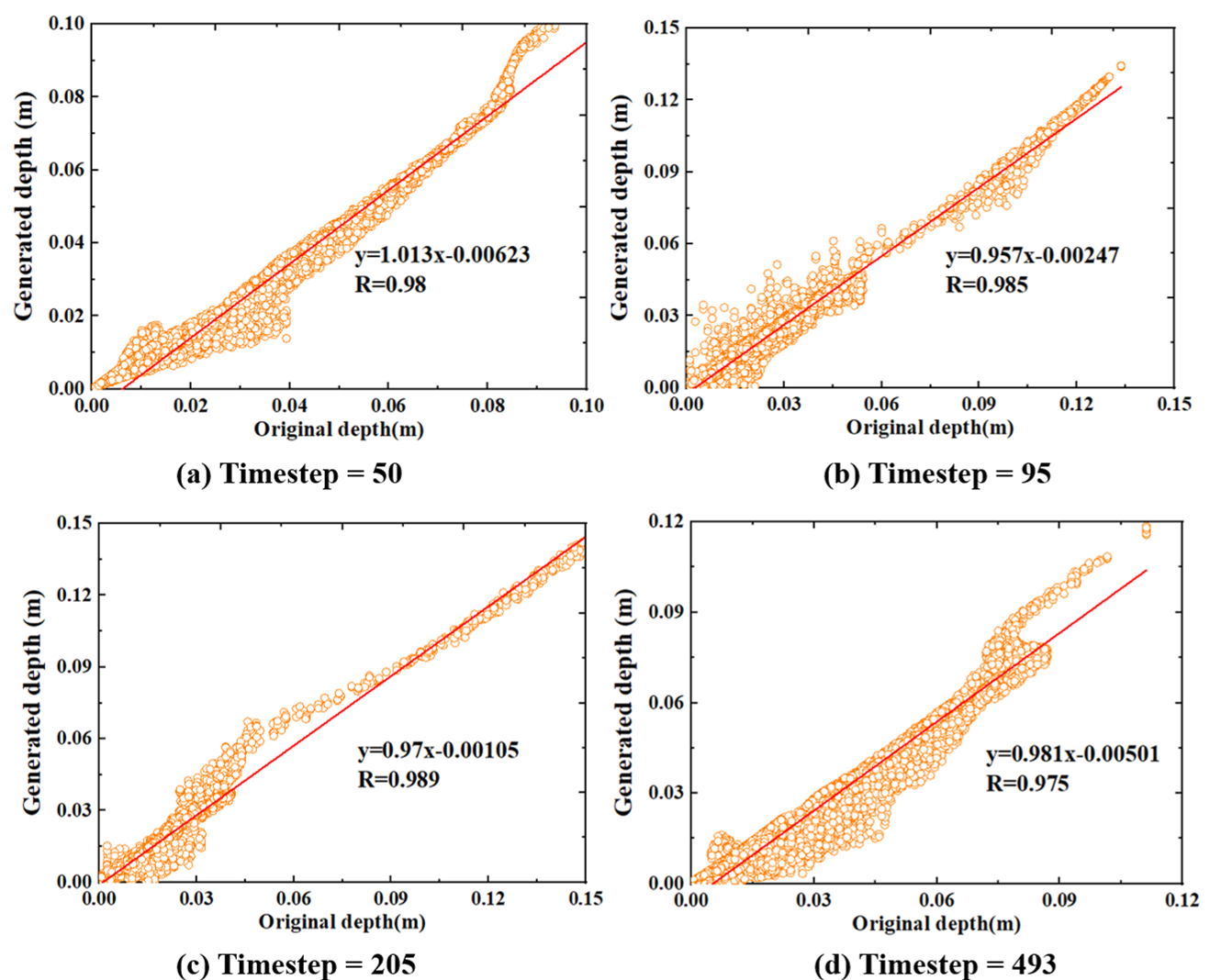}
\caption{The correlation coefficients of water depth solutions between the DCGAN and the original high fidelity model at timesteps  $50, 95, 205, 493$.}
\label{errors_fields2}
\end{figure}

In this case, the numerical illustration of the DCGAN was set up with a series of waves where the wave height varies. For a given input ${\mu \in (\vartheta \backslash {\vartheta _{tr}})}$, \Cref{fields2} shows the comparison of predictive results between the DCGAN and the original high fidelity model. It is observed that most of water depth features are captured compared to the original fluid fields. In addition, the correlation coefficient of results at timesteps $50, 95, 205, 493$ achieves values beyond ${97\%}$ shown in \Cref{errors_fields2}. These results highlight the good performances of the DCGAN in predictive analysis of fluid fields.

\begin{figure}[H]
\centering
\includegraphics[width=0.8\textwidth]{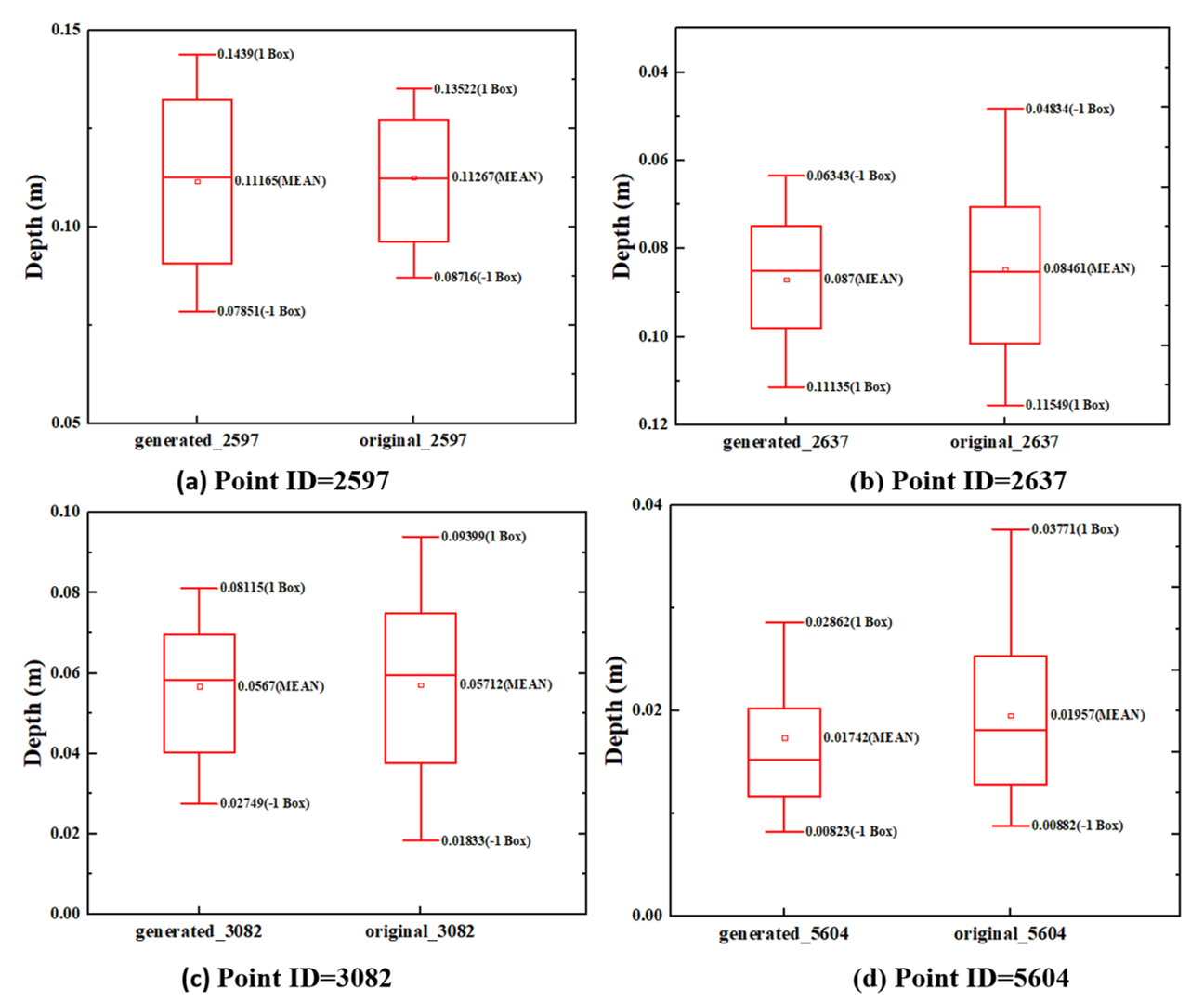}
\caption{Comparison of water depth solutions between the DCGAN and the original high fidelity model at four points during whole simulational period.}
\label{points2}
\end{figure}

The temporal variations of water depth at four points are presented as the box plots in \Cref{points2}. The middle line represents the mean value ${50\%}$, the upper represents the maximum (1 box)  and the lower line denotes the minimum (-1 box). As regards to the mean value, the differences of mean values are less than $0.002 m$ at points ID = 2597, 2637, 3082 and 5604. The depth peak predicted by the DCGAN is $0.1439 m$ at the point ID = 2597, which is comparable to that ($0.1352 m$) from the original high fidelity model. A good performance of the DCGAN in predicting depth peak is achieved at points ID = 2637, 3082, and 5604. As for the minimum depth at the four points, the differences are less than $0.01 m$ as depicted in \Cref{points2}. 

\begin{figure}[H]
\centering
\includegraphics[width=0.7\textwidth]{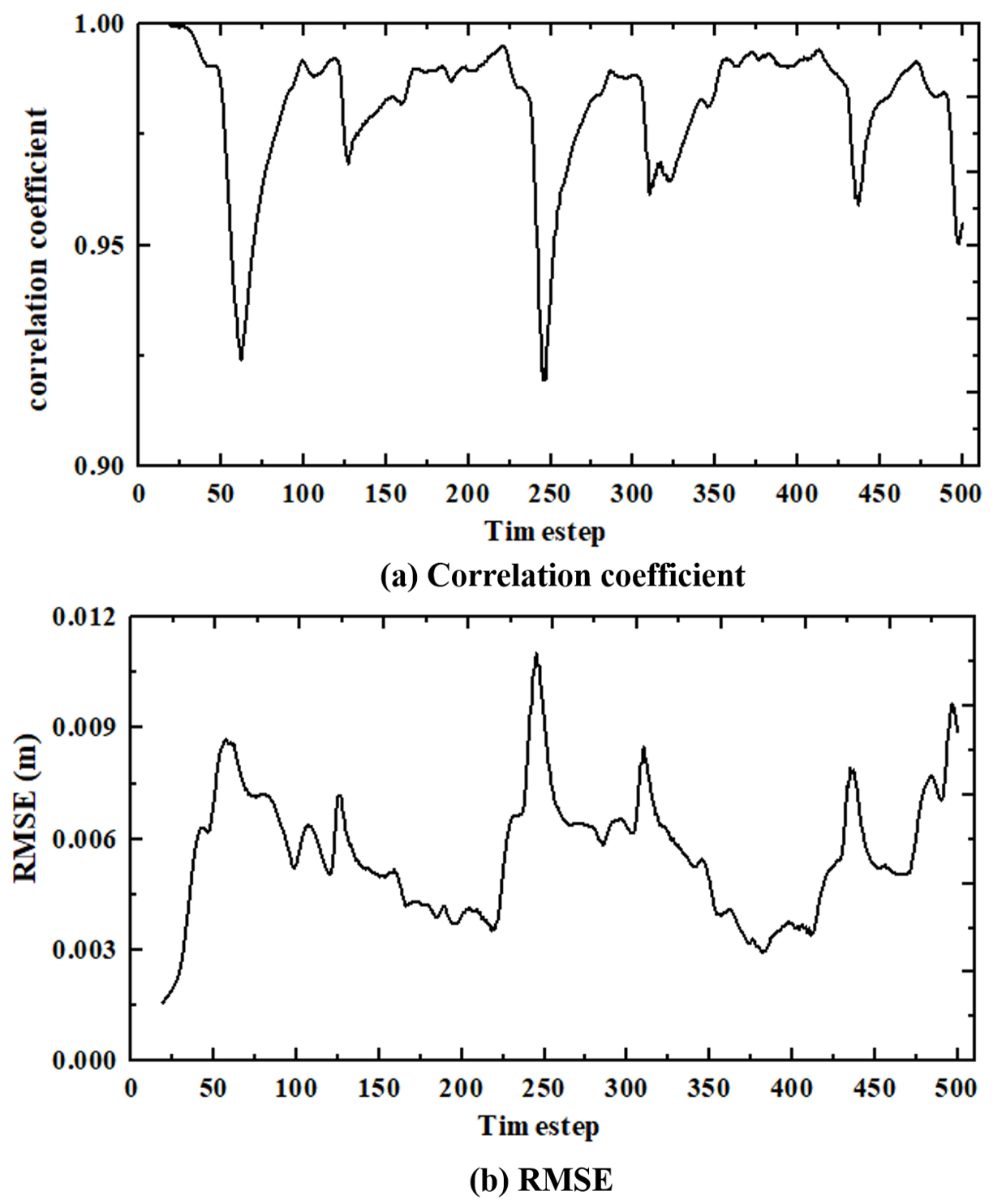}
\caption{The RMSE and correlation coefficients of water depth solutions between the DCGAN and the original high fidelity model during whole simulational period.}
\label{RMSE_R2}
\end{figure}

The error analysis of water depth solutions has been further carried out using the correlation coefficient and RMSE during the simulation period, as shown in \Cref{RMSE_R2}. The values of RMSE fluctuate at different timesteps and are less than $0.012 m$. The correlation coefficient of results between the DCGAN and the high fidelity model exceeds ${90 \%}$.

\subsection{{Computational efficiency}}
The DCGAN and the original high fidelity model were run on a computer (Intel Xeon(R) CPU@ 3.60GHz and a 449.5 GB memory). In terms of computation efficiency, it is worth mentioning that the online CPU cost used for running the DCGAN is less than $0.5 s$, while $5676 s$ for running the original high fidelity model. It can be seen that using the DCGAN, the CPU time is reduced by five orders of magnitude in comparison to the original high fidelity model.

To summarize, the DCGAN has successfully captured fluid flow features and provides the high-resolution fluid fields, especially for the nonlinear fluid flows in time and space. This serves as a technologically important motivation for using the DCGAN to efficiently forecast complicated fluid flows for realistic applications.

\section{Conclusions}
\label{sec7}
In this work, a DCGAN has been proposed for nonlinear spatio-temporal fluid flow prediction. It is the first time that the DCGAN is applied to prediction of spatially-temporally-coherent and high-dimensional nonlinear fluid flows. This novel DCGAN enables rapid and reliable prediction and can be used for providing risk reduction strategies in emergencies. 

The performance of the DCGAN has been illustrated in the case of Hokkaido tsunami in this paper. Two study cases are set up with two series of waves with varying wave phase and wave height. To estimate the accuracy of the DCGAN, a comparison of nonlinear flow results obtained by the DCGAN and the original high fidelity model has been carried out. It is shown that the DCGAN is able to capture the complex flow features and exhibits an overall good agreement with the original high fidelity model. Error analysis has also been performed for further evaluation of the DCGAN through RMSE and correlation coefficient. In comparison to the original high fidelity model, the DCGAN achieves a reasonable accuracy in the whole simulation, while the online CPU cost is reduced by five orders of magnitude.

Overall, a general DCGAN for spatio-temporal fluid flow modelling is presented in this paper. It will be highly promising to apply the above DCGAN to various modelling problems, for example, turbulence modelling, ocean modelling, air pollution, and urban flows. Future work will focus on optimization of uncertainties in models and explore a hybrid machine learning approach (auto-encoder and GAN) for improving the accuracy and efficiency in prediction for complex problems. In this work we focus on the simulation of spatio-temporal distribution during a fixed simulation period for given varied boundary conditions, where the training and validation simulation periods are the same. In future, we will also focus on the real-time forecast-- using the DCGAN for the lead-time prediction of spatio-temporal fluid flows. In addition, for realistic applications, a large number of snapshots (historical datasets), over a long time period, are needed for model training. The Classical Multidimensional Scaling (CMDS) proposed by \citep{kaiser2014cluster, kaiser2017control} can be an efficient method to address this compression issue, which is able to convert a large set of observations into a set of low-dimensional states for visualization. Our future work will further explore the use of other compression methods (e.g. CMDS) along with the GAN to handle high dimensional data for real-time prediction in realistic applications. 

\section*{Acknowledgments}
The Authors acknowledge the support of: China Scholarship Council (No. 201806270238), funding from the EPSRC (MAGIC (EP/N010221/1), INHALE (EP/T003189/1), MUFFINS  (EP/P033148/1) and PREMIERE (EP/T000414/1)), and the Royal Society  (IEC/ NSFC/170563) in the UK. 
\end{spacing}

\renewcommand\refname{References}
\bibliographystyle{model4-names}
\bibliography{paper1}

\begin{thebibliography}{37}
\expandafter\ifx\csname natexlab\endcsname\relax\def\natexlab#1{#1}\fi
\providecommand{\url}[1]{\texttt{#1}}
\providecommand{\href}[2]{#2}
\providecommand{\path}[1]{#1}
\providecommand{\DOIprefix}{doi:}
\providecommand{\ArXivprefix}{arXiv:}
\providecommand{\URLprefix}{URL: }
\providecommand{\Pubmedprefix}{pmid:}
\providecommand{\doi}[1]{\href{http://dx.doi.org/#1}{\path{#1}}}
\providecommand{\Pubmed}[1]{\href{pmid:#1}{\path{#1}}}
\providecommand{\bibinfo}[2]{#2}
\ifx\xfnm\undefined \def\xfnm[#1]{\unskip,\space#1}\fi
\bibitem[{Brunton et~al.(2019)Brunton, Noack and
  Koumoutsakos}]{Brunton2019MachineLF}
\bibinfo{author}{Brunton\xfnm[ S.L.]}, \bibinfo{author}{Noack\xfnm[ B.]},
  \bibinfo{author}{Koumoutsakos\xfnm[ P.]}.
\newblock \bibinfo{title}{Machine learning for fluid mechanics}.
\newblock \bibinfo{journal}{ArXiv}
  \bibinfo{year}{2019};\bibinfo{volume}{abs/1905.11075}.
\bibitem[{Chang and Tsai(2016)}]{chang2016nonlinear}
\bibinfo{author}{Chang\xfnm[ F.J.]}, \bibinfo{author}{Tsai\xfnm[ M.J.]}.
\newblock \bibinfo{title}{A nonlinear spatio-temporal lumping of radar rainfall
  for modeling multi-step-ahead inflow forecasts by data-driven techniques}.
\newblock \bibinfo{journal}{Journal of Hydrology}
  \bibinfo{year}{2016};\bibinfo{volume}{535}:\bibinfo{pages}{256--269}.
\bibitem[{Creswell et~al.(2018)Creswell, White, Dumoulin, Arulkumaran, Sengupta
  and Bharath}]{creswell2018generative}
\bibinfo{author}{Creswell\xfnm[ A.]}, \bibinfo{author}{White\xfnm[ T.]},
  \bibinfo{author}{Dumoulin\xfnm[ V.]}, \bibinfo{author}{Arulkumaran\xfnm[
  K.]}, \bibinfo{author}{Sengupta\xfnm[ B.]}, \bibinfo{author}{Bharath\xfnm[
  A.A.]}.
\newblock \bibinfo{title}{Generative adversarial networks: An overview}.
\newblock \bibinfo{journal}{IEEE Signal Processing Magazine}
  \bibinfo{year}{2018};\bibinfo{volume}{35}(\bibinfo{number}{1}):\bibinfo{pages}{53--65}.
\bibitem[{Fang et~al.(2013)Fang, Pain, Navon, Elsheikh, Du and
  Xiao}]{fang2013non}
\bibinfo{author}{Fang\xfnm[ F.]}, \bibinfo{author}{Pain\xfnm[ C.C.]},
  \bibinfo{author}{Navon\xfnm[ I.]}, \bibinfo{author}{Elsheikh\xfnm[ A.]},
  \bibinfo{author}{Du\xfnm[ J.]}, \bibinfo{author}{Xiao\xfnm[ D.]}.
\newblock \bibinfo{title}{Non-linear petrov--galerkin methods for reduced order
  hyperbolic equations and discontinuous finite element methods}.
\newblock \bibinfo{journal}{Journal of Computational Physics}
  \bibinfo{year}{2013};\bibinfo{volume}{234}:\bibinfo{pages}{540--559}.
\bibitem[{Fang et~al.(2014)Fang, Zhang, Pavlidis, Pain, Buchan and
  Navon}]{fang2014reduced}
\bibinfo{author}{Fang\xfnm[ F.]}, \bibinfo{author}{Zhang\xfnm[ T.]},
  \bibinfo{author}{Pavlidis\xfnm[ D.]}, \bibinfo{author}{Pain\xfnm[ C.]},
  \bibinfo{author}{Buchan\xfnm[ A.]}, \bibinfo{author}{Navon\xfnm[ I.]}.
\newblock \bibinfo{title}{Reduced order modelling of an unstructured mesh air
  pollution model and application in 2d/3d urban street canyons}.
\newblock \bibinfo{journal}{Atmospheric Environment}
  \bibinfo{year}{2014};\bibinfo{volume}{96}:\bibinfo{pages}{96--106}.
\bibitem[{Farimani et~al.(2017)Farimani, Gomes and Pande}]{farimani2017deep}
\bibinfo{author}{Farimani\xfnm[ A.B.]}, \bibinfo{author}{Gomes\xfnm[ J.]},
  \bibinfo{author}{Pande\xfnm[ V.S.]}.
\newblock \bibinfo{title}{Deep learning the physics of transport phenomena}.
\newblock \bibinfo{journal}{arXiv preprint arXiv:170902432}
  \bibinfo{year}{2017};.
\bibitem[{Geneva and Zabaras(2019)}]{geneva2019quantifying}
\bibinfo{author}{Geneva\xfnm[ N.]}, \bibinfo{author}{Zabaras\xfnm[ N.]}.
\newblock \bibinfo{title}{Quantifying model form uncertainty in
  reynolds-averaged turbulence models with bayesian deep neural networks}.
\newblock \bibinfo{journal}{Journal of Computational Physics}
  \bibinfo{year}{2019};.
\bibitem[{Gonzalez and Balajewicz(2018)}]{gonzalez2018learning}
\bibinfo{author}{Gonzalez\xfnm[ F.J.]}, \bibinfo{author}{Balajewicz\xfnm[ M.]}.
\newblock \bibinfo{title}{Learning low-dimensional feature dynamics using deep
  convolutional recurrent autoencoders}.
\newblock \bibinfo{journal}{arXiv preprint arXiv:180801346}
  \bibinfo{year}{2018};.
\bibitem[{Goodfellow(2016)}]{goodfellow2016nips}
\bibinfo{author}{Goodfellow\xfnm[ I.]}.
\newblock \bibinfo{title}{Nips 2016 tutorial: Generative adversarial networks}.
\newblock \bibinfo{journal}{arXiv preprint arXiv:170100160}
  \bibinfo{year}{2016};.
\bibitem[{Goodfellow et~al.(2014)Goodfellow, Pouget-Abadie, Mirza, Xu,
  Warde-Farley, Ozair, Courville and Bengio}]{goodfellow2014generative}
\bibinfo{author}{Goodfellow\xfnm[ I.]}, \bibinfo{author}{Pouget-Abadie\xfnm[
  J.]}, \bibinfo{author}{Mirza\xfnm[ M.]}, \bibinfo{author}{Xu\xfnm[ B.]},
  \bibinfo{author}{Warde-Farley\xfnm[ D.]}, \bibinfo{author}{Ozair\xfnm[ S.]},
  \bibinfo{author}{Courville\xfnm[ A.]}, \bibinfo{author}{Bengio\xfnm[ Y.]}.
\newblock \bibinfo{title}{Generative adversarial nets}.
\newblock In: \bibinfo{booktitle}{Advances in neural information processing
  systems}. \bibinfo{year}{2014}. p. \bibinfo{pages}{2672--2680}.
\bibitem[{Hu et~al.(2019{\natexlab{a}})Hu, Fang, Pain and Navon}]{hu2019rapid}
\bibinfo{author}{Hu\xfnm[ R.]}, \bibinfo{author}{Fang\xfnm[ F.]},
  \bibinfo{author}{Pain\xfnm[ C.]}, \bibinfo{author}{Navon\xfnm[ I.]}.
\newblock \bibinfo{title}{Rapid spatio-temporal flood prediction and
  uncertainty quantification using a deep learning method}.
\newblock \bibinfo{journal}{Journal of Hydrology}
  \bibinfo{year}{2019}{\natexlab{a}};.
\bibitem[{Hu et~al.(2018)Hu, Fang, Salinas and Pain}]{hu2018unstructured}
\bibinfo{author}{Hu\xfnm[ R.]}, \bibinfo{author}{Fang\xfnm[ F.]},
  \bibinfo{author}{Salinas\xfnm[ P.]}, \bibinfo{author}{Pain\xfnm[ C.]}.
\newblock \bibinfo{title}{Unstructured mesh adaptivity for urban flooding
  modelling}.
\newblock \bibinfo{journal}{Journal of hydrology}
  \bibinfo{year}{2018};\bibinfo{volume}{560}:\bibinfo{pages}{354--363}.
\bibitem[{Hu et~al.(2019{\natexlab{b}})Hu, Fang, Salinas, Pain, Domingo and
  Mark}]{hu2019numerical}
\bibinfo{author}{Hu\xfnm[ R.]}, \bibinfo{author}{Fang\xfnm[ F.]},
  \bibinfo{author}{Salinas\xfnm[ P.]}, \bibinfo{author}{Pain\xfnm[ C.]},
  \bibinfo{author}{Domingo\xfnm[ N.S.]}, \bibinfo{author}{Mark\xfnm[ O.]}.
\newblock \bibinfo{title}{Numerical simulation of floods from multiple sources
  using an adaptive anisotropic unstructured mesh method}.
\newblock \bibinfo{journal}{Advances in Water Resources}
  \bibinfo{year}{2019}{\natexlab{b}};\bibinfo{volume}{123}:\bibinfo{pages}{173--188}.
\bibitem[{Isola et~al.(2017)Isola, Zhu, Zhou and Efros}]{isola2017image}
\bibinfo{author}{Isola\xfnm[ P.]}, \bibinfo{author}{Zhu\xfnm[ J.Y.]},
  \bibinfo{author}{Zhou\xfnm[ T.]}, \bibinfo{author}{Efros\xfnm[ A.A.]}.
\newblock \bibinfo{title}{Image-to-image translation with conditional
  adversarial networks}.
\newblock In: \bibinfo{booktitle}{Proceedings of the IEEE conference on
  computer vision and pattern recognition}. \bibinfo{year}{2017}. p.
  \bibinfo{pages}{1125--1134}.
\bibitem[{Kaiser et~al.(2017)Kaiser, Li and Noack}]{kaiser2017control}
\bibinfo{author}{Kaiser\xfnm[ E.]}, \bibinfo{author}{Li\xfnm[ R.]},
  \bibinfo{author}{Noack\xfnm[ B.R.]}.
\newblock \bibinfo{title}{On the control landscape topology}.
\newblock In: \bibinfo{booktitle}{The 20th World Congress of the International
  Federation of Automatic Control (IFAC)}. \bibinfo{year}{2017}. p.
  \bibinfo{pages}{1--5}.
\bibitem[{Kaiser et~al.(2014)Kaiser, Noack, Cordier, Spohn, Segond, Abel,
  Daviller, {\"O}sth, Krajnovi{\'c} and Niven}]{kaiser2014cluster}
\bibinfo{author}{Kaiser\xfnm[ E.]}, \bibinfo{author}{Noack\xfnm[ B.R.]},
  \bibinfo{author}{Cordier\xfnm[ L.]}, \bibinfo{author}{Spohn\xfnm[ A.]},
  \bibinfo{author}{Segond\xfnm[ M.]}, \bibinfo{author}{Abel\xfnm[ M.]},
  \bibinfo{author}{Daviller\xfnm[ G.]}, \bibinfo{author}{{\"O}sth\xfnm[ J.]},
  \bibinfo{author}{Krajnovi{\'c}\xfnm[ S.]}, \bibinfo{author}{Niven\xfnm[
  R.K.]}.
\newblock \bibinfo{title}{Cluster-based reduced-order modelling of a mixing
  layer}.
\newblock \bibinfo{journal}{Journal of Fluid Mechanics}
  \bibinfo{year}{2014};\bibinfo{volume}{754}:\bibinfo{pages}{365--414}.
\bibitem[{Kalteh(2013)}]{kalteh2013monthly}
\bibinfo{author}{Kalteh\xfnm[ A.M.]}.
\newblock \bibinfo{title}{Monthly river flow forecasting using artificial
  neural network and support vector regression models coupled with wavelet
  transform}.
\newblock \bibinfo{journal}{Computers \& Geosciences}
  \bibinfo{year}{2013};\bibinfo{volume}{54}:\bibinfo{pages}{1--8}.
\bibitem[{Kasiviswanathan et~al.(2016)Kasiviswanathan, He, Sudheer and
  Tay}]{kasiviswanathan2016potential}
\bibinfo{author}{Kasiviswanathan\xfnm[ K.]}, \bibinfo{author}{He\xfnm[ J.]},
  \bibinfo{author}{Sudheer\xfnm[ K.]}, \bibinfo{author}{Tay\xfnm[ J.H.]}.
\newblock \bibinfo{title}{Potential application of wavelet neural network
  ensemble to forecast streamflow for flood management}.
\newblock \bibinfo{journal}{Journal of hydrology}
  \bibinfo{year}{2016};\bibinfo{volume}{536}:\bibinfo{pages}{161--173}.
\bibitem[{Kisi et~al.(2012)Kisi, Shiri and Nikoofar}]{kisi2012forecasting}
\bibinfo{author}{Kisi\xfnm[ O.]}, \bibinfo{author}{Shiri\xfnm[ J.]},
  \bibinfo{author}{Nikoofar\xfnm[ B.]}.
\newblock \bibinfo{title}{Forecasting daily lake levels using artificial
  intelligence approaches}.
\newblock \bibinfo{journal}{Computers \& Geosciences}
  \bibinfo{year}{2012};\bibinfo{volume}{41}:\bibinfo{pages}{169--180}.
\bibitem[{Laloy et~al.(2017)Laloy, H{\'e}rault, Jacques and
  Linde}]{laloy2017efficient}
\bibinfo{author}{Laloy\xfnm[ E.]}, \bibinfo{author}{H{\'e}rault\xfnm[ R.]},
  \bibinfo{author}{Jacques\xfnm[ D.]}, \bibinfo{author}{Linde\xfnm[ N.]}.
\newblock \bibinfo{title}{Efficient training-image based geostatistical
  simulation and inversion using a spatial generative adversarial neural
  network}.
\newblock \bibinfo{journal}{arXiv preprint arXiv:170804975}
  \bibinfo{year}{2017};.
\bibitem[{Ledig et~al.(2017)Ledig, Theis, Husz{\'a}r, Caballero, Cunningham,
  Acosta, Aitken, Tejani, Totz, Wang et~al.}]{ledig2017photo}
\bibinfo{author}{Ledig\xfnm[ C.]}, \bibinfo{author}{Theis\xfnm[ L.]},
  \bibinfo{author}{Husz{\'a}r\xfnm[ F.]}, \bibinfo{author}{Caballero\xfnm[
  J.]}, \bibinfo{author}{Cunningham\xfnm[ A.]}, \bibinfo{author}{Acosta\xfnm[
  A.]}, \bibinfo{author}{Aitken\xfnm[ A.]}, \bibinfo{author}{Tejani\xfnm[ A.]},
  \bibinfo{author}{Totz\xfnm[ J.]}, \bibinfo{author}{Wang\xfnm[ Z.]}, et~al.
\newblock \bibinfo{title}{Photo-realistic single image super-resolution using a
  generative adversarial network}.
\newblock In: \bibinfo{booktitle}{Proceedings of the IEEE conference on
  computer vision and pattern recognition}. \bibinfo{year}{2017}. p.
  \bibinfo{pages}{4681--4690}.
\bibitem[{Li and Wand(2016)}]{li2016precomputed}
\bibinfo{author}{Li\xfnm[ C.]}, \bibinfo{author}{Wand\xfnm[ M.]}.
\newblock \bibinfo{title}{Precomputed real-time texture synthesis with
  markovian generative adversarial networks}.
\newblock In: \bibinfo{booktitle}{European Conference on Computer Vision}.
  \bibinfo{organization}{Springer}; \bibinfo{year}{2016}. p.
  \bibinfo{pages}{702--716}.
\bibitem[{Lohani et~al.(2014)Lohani, Goel and Bhatia}]{lohani2014improving}
\bibinfo{author}{Lohani\xfnm[ A.K.]}, \bibinfo{author}{Goel\xfnm[ N.]},
  \bibinfo{author}{Bhatia\xfnm[ K.]}.
\newblock \bibinfo{title}{Improving real time flood forecasting using fuzzy
  inference system}.
\newblock \bibinfo{journal}{Journal of hydrology}
  \bibinfo{year}{2014};\bibinfo{volume}{509}:\bibinfo{pages}{25--41}.
\bibitem[{Miyanawala and Jaiman(2017)}]{miyanawala2017efficient}
\bibinfo{author}{Miyanawala\xfnm[ T.P.]}, \bibinfo{author}{Jaiman\xfnm[ R.K.]}.
\newblock \bibinfo{title}{An efficient deep learning technique for the
  navier-stokes equations: Application to unsteady wake flow dynamics}.
\newblock \bibinfo{journal}{arXiv preprint arXiv:171009099}
  \bibinfo{year}{2017};.
\bibitem[{Nair et~al.(2019)Nair, Yeh, Kaiser, Noack, Brunton and
  Taira}]{nair2019cluster}
\bibinfo{author}{Nair\xfnm[ A.G.]}, \bibinfo{author}{Yeh\xfnm[ C.A.]},
  \bibinfo{author}{Kaiser\xfnm[ E.]}, \bibinfo{author}{Noack\xfnm[ B.R.]},
  \bibinfo{author}{Brunton\xfnm[ S.L.]}, \bibinfo{author}{Taira\xfnm[ K.]}.
\newblock \bibinfo{title}{Cluster-based feedback control of turbulent
  post-stall separated flows}.
\newblock \bibinfo{journal}{Journal of Fluid Mechanics}
  \bibinfo{year}{2019};\bibinfo{volume}{875}:\bibinfo{pages}{345--375}.
\bibitem[{{\"O}sth et~al.(2015){\"O}sth, Kaiser, Krajnovi{\'c} and
  Noack}]{osth2015cluster}
\bibinfo{author}{{\"O}sth\xfnm[ J.]}, \bibinfo{author}{Kaiser\xfnm[ E.]},
  \bibinfo{author}{Krajnovi{\'c}\xfnm[ S.]}, \bibinfo{author}{Noack\xfnm[
  B.R.]}.
\newblock \bibinfo{title}{Cluster-based reduced-order modelling of the flow in
  the wake of a high speed train}.
\newblock \bibinfo{journal}{Journal of Wind Engineering and Industrial
  Aerodynamics}
  \bibinfo{year}{2015};\bibinfo{volume}{145}:\bibinfo{pages}{327--338}.
\bibitem[{Pain et~al.(2001)Pain, Umpleby, De~Oliveira and
  Goddard}]{pain2001tetrahedral}
\bibinfo{author}{Pain\xfnm[ C.]}, \bibinfo{author}{Umpleby\xfnm[ A.]},
  \bibinfo{author}{De~Oliveira\xfnm[ C.]}, \bibinfo{author}{Goddard\xfnm[ A.]}.
\newblock \bibinfo{title}{Tetrahedral mesh optimisation and adaptivity for
  steady-state and transient finite element calculations}.
\newblock \bibinfo{journal}{Computer Methods in Applied Mechanics and
  Engineering}
  \bibinfo{year}{2001};\bibinfo{volume}{190}(\bibinfo{number}{29-30}):\bibinfo{pages}{3771--3796}.
\bibitem[{Prakash et~al.(2014)Prakash, Sudheer and
  Srinivasan}]{prakash2014improved}
\bibinfo{author}{Prakash\xfnm[ O.]}, \bibinfo{author}{Sudheer\xfnm[ K.]},
  \bibinfo{author}{Srinivasan\xfnm[ K.]}.
\newblock \bibinfo{title}{Improved higher lead time river flow forecasts using
  sequential neural network with error updating}.
\newblock \bibinfo{journal}{Journal of Hydrology and Hydromechanics}
  \bibinfo{year}{2014};\bibinfo{volume}{62}(\bibinfo{number}{1}):\bibinfo{pages}{60--74}.
\bibitem[{Radford et~al.(2015)Radford, Metz and
  Chintala}]{radford2015unsupervised}
\bibinfo{author}{Radford\xfnm[ A.]}, \bibinfo{author}{Metz\xfnm[ L.]},
  \bibinfo{author}{Chintala\xfnm[ S.]}.
\newblock \bibinfo{title}{Unsupervised representation learning with deep
  convolutional generative adversarial networks}.
\newblock \bibinfo{journal}{arXiv preprint arXiv:151106434}
  \bibinfo{year}{2015};.
\bibitem[{Reed et~al.(2016)Reed, Akata, Yan, Logeswaran, Schiele and
  Lee}]{reed2016generative}
\bibinfo{author}{Reed\xfnm[ S.]}, \bibinfo{author}{Akata\xfnm[ Z.]},
  \bibinfo{author}{Yan\xfnm[ X.]}, \bibinfo{author}{Logeswaran\xfnm[ L.]},
  \bibinfo{author}{Schiele\xfnm[ B.]}, \bibinfo{author}{Lee\xfnm[ H.]}.
\newblock \bibinfo{title}{Generative adversarial text to image synthesis}.
\newblock \bibinfo{journal}{arXiv preprint arXiv:160505396}
  \bibinfo{year}{2016};.
\bibitem[{Reichstein et~al.(2019)Reichstein, Camps-Valls, Stevens, Jung,
  Denzler, Carvalhais et~al.}]{reichstein2019deep}
\bibinfo{author}{Reichstein\xfnm[ M.]}, \bibinfo{author}{Camps-Valls\xfnm[
  G.]}, \bibinfo{author}{Stevens\xfnm[ B.]}, \bibinfo{author}{Jung\xfnm[ M.]},
  \bibinfo{author}{Denzler\xfnm[ J.]}, \bibinfo{author}{Carvalhais\xfnm[ N.]},
  et~al.
\newblock \bibinfo{title}{Deep learning and process understanding for
  data-driven earth system science}.
\newblock \bibinfo{journal}{Nature}
  \bibinfo{year}{2019};\bibinfo{volume}{566}(\bibinfo{number}{7743}):\bibinfo{pages}{195}.
\bibitem[{Str{\"o}fer et~al.(2018)Str{\"o}fer, Wu, Xiao and
  Paterson}]{strofer2018data}
\bibinfo{author}{Str{\"o}fer\xfnm[ C.M.]}, \bibinfo{author}{Wu\xfnm[ J.]},
  \bibinfo{author}{Xiao\xfnm[ H.]}, \bibinfo{author}{Paterson\xfnm[ E.]}.
\newblock \bibinfo{title}{Data-driven, physics-based feature extraction from
  fluid flow fields}.
\newblock \bibinfo{journal}{arXiv preprint arXiv:180200775}
  \bibinfo{year}{2018};.
\bibitem[{Wang et~al.(2018)Wang, Xiao, Fang, Govindan, Pain and
  Guo}]{wang2018model}
\bibinfo{author}{Wang\xfnm[ Z.]}, \bibinfo{author}{Xiao\xfnm[ D.]},
  \bibinfo{author}{Fang\xfnm[ F.]}, \bibinfo{author}{Govindan\xfnm[ R.]},
  \bibinfo{author}{Pain\xfnm[ C.C.]}, \bibinfo{author}{Guo\xfnm[ Y.]}.
\newblock \bibinfo{title}{Model identification of reduced order fluid dynamics
  systems using deep learning}.
\newblock \bibinfo{journal}{International Journal for Numerical Methods in
  Fluids}
  \bibinfo{year}{2018};\bibinfo{volume}{86}(\bibinfo{number}{4}):\bibinfo{pages}{255--268}.
\bibitem[{Xiao et~al.(2019)Xiao, Fang, Zheng, Pain and Navon}]{xiao2019machine}
\bibinfo{author}{Xiao\xfnm[ D.]}, \bibinfo{author}{Fang\xfnm[ F.]},
  \bibinfo{author}{Zheng\xfnm[ J.]}, \bibinfo{author}{Pain\xfnm[ C.]},
  \bibinfo{author}{Navon\xfnm[ I.]}.
\newblock \bibinfo{title}{Machine learning-based rapid response tools for
  regional air pollution modelling}.
\newblock \bibinfo{journal}{Atmospheric Environment}
  \bibinfo{year}{2019};\bibinfo{volume}{199}:\bibinfo{pages}{463--473}.
\bibitem[{Xiao et~al.(2016)}]{xiao2016non}
\bibinfo{author}{Xiao\xfnm[ D.]}, et~al.
\newblock \bibinfo{title}{Non-intrusive reduced order models and their
  applications}.
\newblock Ph.D. thesis; Imperial College London; \bibinfo{year}{2016}.
\bibitem[{Xie et~al.(2018)Xie, Franz, Chu and Thuerey}]{xie2018tempogan}
\bibinfo{author}{Xie\xfnm[ Y.]}, \bibinfo{author}{Franz\xfnm[ E.]},
  \bibinfo{author}{Chu\xfnm[ M.]}, \bibinfo{author}{Thuerey\xfnm[ N.]}.
\newblock \bibinfo{title}{tempogan: A temporally coherent, volumetric gan for
  super-resolution fluid flow}.
\newblock \bibinfo{journal}{ACM Transactions on Graphics (TOG)}
  \bibinfo{year}{2018};\bibinfo{volume}{37}(\bibinfo{number}{4}):\bibinfo{pages}{95}.
\bibitem[{Yang et~al.(2017)Yang, Asanjan, Welles, Gao, Sorooshian and
  Liu}]{yang2017developing}
\bibinfo{author}{Yang\xfnm[ T.]}, \bibinfo{author}{Asanjan\xfnm[ A.A.]},
  \bibinfo{author}{Welles\xfnm[ E.]}, \bibinfo{author}{Gao\xfnm[ X.]},
  \bibinfo{author}{Sorooshian\xfnm[ S.]}, \bibinfo{author}{Liu\xfnm[ X.]}.
\newblock \bibinfo{title}{Developing reservoir monthly inflow forecasts using
  artificial intelligence and climate phenomenon information}.
\newblock \bibinfo{journal}{Water Resources Research}
  \bibinfo{year}{2017};\bibinfo{volume}{53}(\bibinfo{number}{4}):\bibinfo{pages}{2786--2812}.

\end{thebibliography}

\end{document}